\documentclass{vgtc}                 

\ifpdf
  \pdfoutput=1\relax                   
  \usepackage{graphicx}                
  \DeclareGraphicsExtensions{.pdf,.png,.jpg,.jpeg} 
\else
  \usepackage{graphicx}                
  \DeclareGraphicsExtensions{.eps}     
\fi%

\graphicspath{{figures/}{pictures/}{images/}{./}} 

\usepackage{microtype}                 
\PassOptionsToPackage{warn}{textcomp}  
\usepackage{textcomp}                  
\usepackage{mathptmx}                  
\usepackage{times}                     
\usepackage{cite}                      
\usepackage{tabu}                      
\usepackage{booktabs}                  

\usepackage{enumitem}
\usepackage{amsmath}
\usepackage{amssymb}
\usepackage{bm}
\usepackage{float}

\onlineid{8347}

\vgtccategory{Research}

\vgtcinsertpkg

\title{Exploring the Effects of VR Activities on Stress Relief: A Comparison of Sitting-in-Silence, VR Meditation, and VR Smash Room}

\author{Dongyun Han\thanks{e-mail: dongyun.han@usu.edu}\\ %
        \scriptsize Utah State University %
\and Donghoon Kim\thanks{e-mail: donghoon.kim@usu.edu}\\ %
     \scriptsize Utah State University %
\and Kangsoo Kim\thanks{e-mail: kangsoo.kim@ucalgary.ca}\\ %
     \scriptsize University of Calgary %
\and Isaac Cho\thanks{e-mail: isaac.cho@usu.edu}\\ %
     {\scriptsize Utah State University }}

\teaser{
  \centering
  \includegraphics[width=\linewidth]{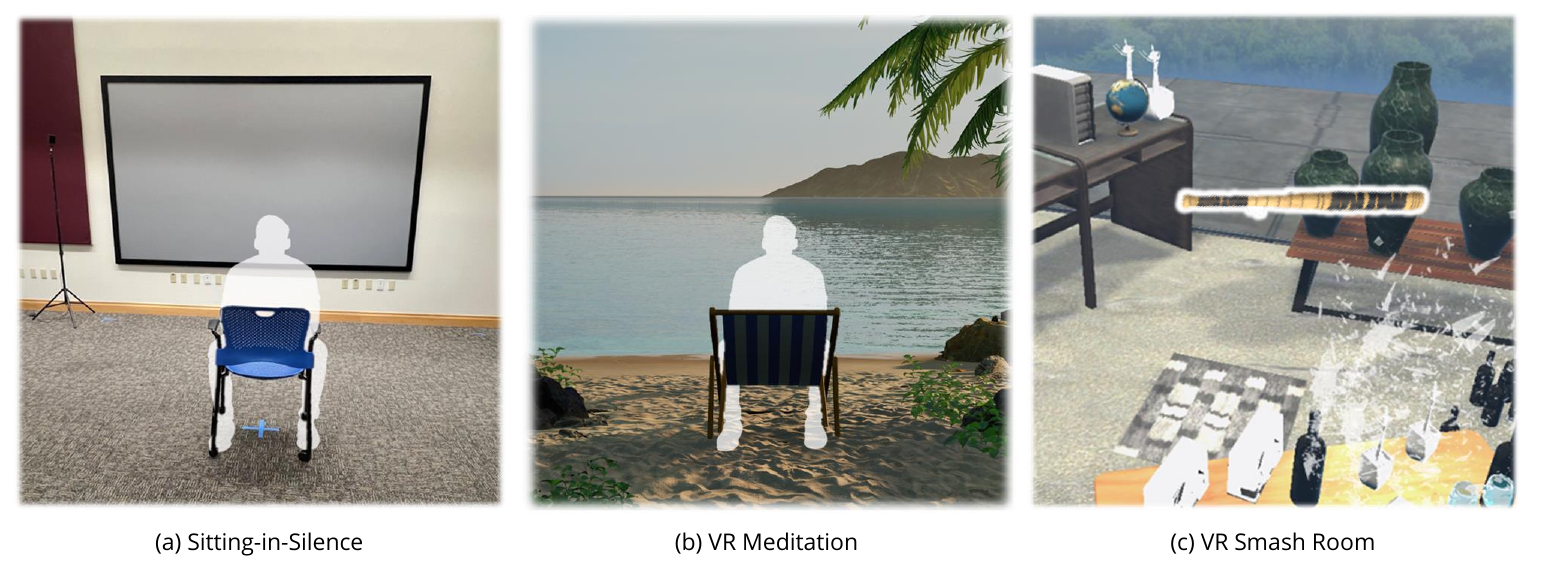}
  \caption{Our study evaluated the effects of one traditional (real-world) and two VR treatments on stress relief. (a) Sitting-in-Silence is a mindfulness observation method asking users to sit in a chair and relax with no guided voice in reality. (b) VR Meditation presents an immersive beach. Users can take a rest by sitting in a chair or exploring the scene. (c) VR Smash Room is designed for users to do the physical activity by smashing virtual objects. The object-breaking experience incorporates crashing sounds and visual effects.}
  \label{fig:teaser}
}

\abstract{
In our lives, we encounter various stressors that may cause negative mental and bodily reactions to make us feel frustrated, angry, or irritated. Effective methods to manage or reduce stress and anxiety are essential for a healthy life, and several stress-management approaches are found to be useful for stress relief, such as meditation, taking a rest, walking around nature, or even breaking things in a smash room. 
Previous research has revealed that certain experiences in virtual reality (VR) are effective for reducing stress as traditional real-world methods. 
However, it is still unclear how the stress relief effects are associated with other factors like individual user profile in terms of different treatment activities. 
In this paper, we report our findings from a formal user study that investigates the effects of two virtual activities: (1) VR Meditation and (2) VR Smash Room experience, compared with a traditional Sitting-in-Silence method. 
Our results show that VR Meditation has a better stress relief effect compared to VR Smash Room and Sitting-in-Silence, and the effects of the treatments are correlated with the participants' personalities. We discuss the findings and implications addressing potential benefits/impacts of different stress-relief activities in VR.
} 

\CCScatlist{
  \CCScatTwelve{Human-centered computing}{Human computer interaction (HCI)}{Empirical studies in HCI}{};
  \CCScatTwelve{Computing methodologies}{Computer graphics}{Graphics systems and interfaces}{Virtual reality}
}

\begin{document}


\firstsection{Introduction}

\maketitle
\label{sec_introduction}
Nowadays, people encounter various stressors, such as anxiety-inducing tasks, work deadlines, or discomfort in social situations. 
These stressors may cause negative mental and bodily reactions to make people feel frustrated, angry, or irritated~\cite{MONROE2016109}. 
Different stress and anxiety management approaches are found to be useful for stress and anxiety relief~\cite{Lehrer1994}. 
For example, \emph{meditation}, a mental practice that can be performed in different ways (e.g., mindfulness), is a well-known stress management method~\cite{Oman2008}.
Another example is a \emph{Smash Room}, also known as a rage or anger room, where people can destroy objects to reduce their stress~\cite{Bennett2017}.
While there is an increasing public concern about possible resulting diseases, such as heart attacks and depression, more studies to explore the impacts of different stress-relief methods are required and emphasized.

For decades, Virtual Reality (VR) has been widely used in the context of mental (and physical) therapy given the potential benefits as a psychotherapy treatment tool~\cite{park2019literature}, including anxiety management~\cite{mantovani2003virtual}, acrophobia treatment~\cite{opdyke1995effectiveness}, and post-traumatic stress disorder (PTSD) therapy~\cite{mistry2020meditating, reger2019does}. 
While addressing unique advantages in the use of VR, such as safety, fun, reproducibility, and realism in the experience, previous research has revealed that certain experiences in VR could provide positive effects on stress relief as traditional methods do in reality---possibly even more effective in some cases.
For instance, Tarrant et al.~\cite{tarrant2022feasibility} and Kaplan-Rakowski et al.~\cite{kaplan2021impact} investigated the effect of VR-based meditation by comparing it to conventional audio- and video-based meditation methods.
They reported that VR meditation provided better or comparable experiences to traditional meditation methods. 
They attributed this to the fact that immersive experience in VR allows users to rehearse mindfulness skills, resolve environmental distractions, and divert their attention from the real world.
Thanks to the recent advances and accessibility of off-the-shelf immersive VR devices and technology, the use of VR is considered for the general public to manage their stress and anxiety levels in their daily lives.
However, there is still a dearth of research that investigates the effects of different stress relief methods in VR on stress reduction as well as the potential relationship with the user profile, such as personality. 

Addressing the knowledge gaps aforementioned, we establish the following research questions (RQs):
\begin{itemize}[leftmargin=0.9cm]
    \item [\textbf{RQ1}.] How effective are stress-management methods in VR for stress reliefduction compared to a traditional method in reality (e.g., resting)?
    \item [\textbf{RQ2}.] How are individual profile factors (e.g., personality traits or tendencies) associated with the effectiveness of stress-management methods?
\end{itemize}

To tackle these questions, we conduct a formal user study with 40 participants to evaluate two stress-management methods in immersive VR environments: 1) \emph{VR Meditation} and 2) \emph{VR Smash Room} experiences, while comparing them with \emph{Sitting-in-Silence} in the real world.  
In the study, participants perform a cognitively challenging task to induce the increase of stress before experiencing the VR stress-management methods and the Sitting-in-Silence, and they report their levels of stress/anxiety before and after the experiences.
While our results show that all three methods (VR Meditation, VR Smash Room, and Sitting-in-Silence) are effective in reducing the level of stress, the meditation experience in VR is more effective than the other two methods.
The results also reveal that there are substantial or moderate associations between the stress-relief effects of VR activities and the participants' personality traits.
Our findings are different from what we initially expected based on previous research that showed positive effects of physical activities in VR on relieving fatigue and reducing depression~\cite{jones2017real}.
We justify our results and findings while discussing possible influences and implications of different VR activities for stress reduction.


\section{Related Work}
\label{Sec:RelatedWork}

\subsection{Stress Management and Measures}

Stress is a sense of emotional or physical tension, and is one of the dominant reasons for physical and mental disorders and diseases, such as heart disease, depression, and asthma~\cite{Yaribeygi2017}.
Previous studies have introduced many effective stress management strategies like mental or physical activities~\cite{Mariotti2015,Cohen2007,Salleh2008}. 
They have also demonstrated that the strategies are advantageous not only for clinical patients but also for the non-clinical general population, such as normal workers and students~\cite{irving2009cultivating, montero2015mindfulness}.
Such strategies have been shown to improve immune function~\cite{grossman2004mindfulness}, working memory~\cite{chambers2008impact, jha2010examining}, and well-being~\cite{campos2016meditation, baer2008construct}.

Mental activities, also known as meditation, not only include mindfulness observation and visual/auditory attention but sometimes even involve physical activities like yoga and Tai Chi~\cite{Matko2019}. 
These activities have different effects depending on their stimuli type, activity time, and instructor existence~\cite{ospina2007meditation, sedlmeier2018psychological}. 
In particular, when physical activities are involved, there would be different impacts on the human body and mind with respect to intensities, types, and frequency of exercise, which in turn could influence the effectiveness in stress reduction~\cite{Herbert2020,Filaire1996}. 
Previous studies reported that even short and temporary exercise or a combination of different activities has significantly reduced stress.
For example, Matzer et al.~\cite{matzer2018combining} examined the effects of combining walking activity and relaxation and reported that such a combination of activities is more effective for stress relief than only walking or resting.

As an example of physical activity, Smash Room allows users to freely break and destroy things in a room for stress relief~\cite{Bennett2017}. 
Despite its increasing popularity, there is still a question on the actual effects of object-breaking activities for stress relief with safety concerns.  
A clinical psychologist with expertise in stress assessment and management pointed out a lack of evidence of the Smash Room effects~\cite{Martin2016}. 

To study the effects of different stress-management approaches, self-reported surveys for subjective perception measures~\cite{Brown2012,Lovibond1993,Spielberger1971} and bio-markers for objective physical/physiological measures~\cite{Dorsey2022} are widely used to evaluate the stress level reliably and accurately. Some earlier studies investigated the association between these subjective and objective measures and revealed a certain degree of association~\cite{weckesser2019psychometric}.

\subsection{Effects of Stress-Relief Treatments in VR and Reality}

With the advancement of technology, various devices have been utilized in order to aid in stress management. 
For instance, mobile devices like smartphones and tablets have made audio- and video-based meditations easily accessible, and using such devices has been shown to be effective for stress management~\cite{flett2019mobile, yang2018happier}.
As VR devices are getting lighter and cheaper, immersive virtual simulations have been tried to treat mental illnesses in the field of clinical domains, such as PTSD~\cite{aiken2015posttraumatic, difede2014d, menelas2018use}, anxiety disorder~\cite{mccann2014virtual}, phobia~\cite{gebara2015virtual} and fear of heights~\cite{freeman2018automated, opdyke1995effectiveness}. 
Regarding the practicality, VR interventions have advantages in two folds: first, they could be less expensive in terms of both space and price, and second, they provide a level of ubiquity and variability that are not typically attainable with traditional methods~\cite{vaquero2021virtual}.

Previous studies investigated the effects of VR-assisted meditations on stress reduction while comparing them to traditional methods in the real world.
Tarrant et al.~\cite{tarrant2018virtual} reported that a VR-based mindfulness intervention significantly improved outcomes in reducing depression and anxiety than an audio-based guided meditation.
They used self-reported anxiety symptoms (e.g., STAI) and electroencephalogram (EEG) from anxious participants.
The participants were asked to practice mindfulness observation in a photo-realistic vast plain generated from a 360-degree video. 
They concluded that the stress-relief effects of the VR intervention are consistent with a physiological reduction of anxiety.
Kaplan-Rakowski et al.~\cite{kaplan2021impact} also claimed that VR could be a potential medium for students to be calmed effectively before stressful events like exams.
This is due to the potential of VR to foster imagination processes by providing an immersive environment, which cannot be easily supported by traditional methods.

However, findings about the impacts of VR experiences on stress relief are not always consistent but sometimes controversial. 
According to other studies, activities in VR might not be significantly effective compared to traditional methods.
Waller et al.~\cite{waller2021meditating} compared a 360-degree video-based VR approach to audio-based meditation.
They reported no great differences between the methods although participants felt a high level of presence in the VR condition. 
Vaquero-Blasco et al.~\cite{vaquero2020virtual} also reported that VR applications mimicking chromotherapy had the same stress relief effects as being in a real-world environment.
Kim et al.~\cite{kim2021effect} used both self-evaluation and biomarkers to compare the effects of VR experience and traditional biofeedback methods on stress relief.
In their study, both VR and biofeedback methods decreased the self-reported stress level, but the effects were not significantly different among the methods.
Interestingly, however, the objective biofeedback showed significant differences between the VR and biofeedback methods.

\subsection{Virtual Settings and Interventions for Stress-Relief}

Immersive experiences from virtual interventions could address challenges to environmental distraction by providing visual and auditory feedback, shifting attention away from the real-world environment~\cite{matzer2018combining, dooris2013expert, berto2014role}. 
Zainudin et al.~\cite{zainudin2014stress}, for example, conducted a user study to compare voice-guided and non-guided meditation exercises in a VR setting. 
They found that the voice-guided exercise was more helpful than the non-guided setting to improve positive feelings, such as confidence, happiness, and pride. On the other hand, they reported that the non-guided setting was more effective in reducing the negative emotions of participants. 

Most VR approaches presented natural environments for users to practice mindfulness observation~\cite{yildirim2020efficacy, li2020effects, chandrasiri2020virtual}.
Zaharuddin et al.~\cite{zaharuddin2019factors} investigated which environmental factors in VR play a positive role in users' relaxation. 
They found five attraction factors: visual attractiveness (e.g., animal or water movement~\cite{nordh2009components, nordh2013pocket}), preferences (e.g., realism, sound, etc), user comfort, navigation method, and interaction with environments.
These factors are taken into account for our study and discussed in Section~\ref{Sec:StudyDesign}.

Physical exercise in VR is also being recognized as a novel experience for healthy lives.
According to earlier studies~\cite{qian2020effectiveness, meyns2017effect}, VR exercise could effectively reduce fatigue and depression, whether users are healthy or injured (e.g., rehabilitation purpose). 
Persson et al.~\cite{persson2021virtual} used a VR Smash Room as a rehabilitation method for cancer patients.
They found that the destructive activities could bring the patients positive pleasure, but warned that VR sickness could have a negative impact on them.
The aforementioned studies have investigated the effects of single-stimulus VR activities (e.g., exercise or meditation in VR).
They have consistently reported that VR activities have similar or better effects than conventional approaches.
However, it still remains unclear which VR activity between mental and physical activities is more effective for users to relieve their stress.
In this study, we evaluate the stress-relief effects of two VR treatments compared to one traditional real-world treatment.

\section{Experiment}
\label{Sec:Experiment}

This section describes the details of our study that investigates the effects of different stress relief treatments in VR and the real world.

\subsection{Participants}
We recruited a total of 45 participants through the university's student participants recruitment system, SONA.
We excluded five participants because their stress level was not increased by our stress-inducing task, which is an inclusion criterion for our study.
Therefore, data collected from 40 out of 45 participants were analyzed in this work (13 male, 27 female; the average age is 19, ranging from 18 to 22). 
All the participants had 20/20 (or corrected 20/20) vision and had no impairments in the use of VR devices. 
The study took approximately one hour, and the participants were rewarded with 1 SONA credit as stipulated based on the SONA policy.

\subsection{Study Design}
\label{Sec:StudyDesign}

Each participant experienced only one of the three stress-relief conditions: Sitting-in-Silence, VR Meditation, and VR Smash Room. The choice of Sitting-in-Silence and VR Meditation was influenced by mindfulness observation methods in the real and virtual worlds, respectively. Meanwhile, the VR Smash Room condition was chosen to represent a physical activity condition in VR.

\paragraph{Sitting-in-Silence:} 
The Sitting-in-Silence condition involves resting in a quiet space and provides an opportunity for mindfulness observation. 
Mindfulness is defined as ``the act of consciously focusing the mind in the present moment without judgment and attachment to the moment~\cite{linehan2014dbt}.'' It is known to promote good mental and physical health. 
For our study, a non-guided approach (no instruction voice) was considered as Zainudin et al.~\cite{zainudin2014stress} suggested. 
The participants were asked to sit on a chair and relax for 5 minutes as shown in Fig.~\ref{fig:teaser}(a).
In this condition, the participants were not forced to close their eyes, and they were not permitted to use any electric devices like their smartphones and smartwatches.
The Sitting-in-Silence group includes 13 participants (1 male, 12 female) in this study. 11 of these participants had prior experience with VR devices, and on average, they rated their familiarity with VR at 3.2 out of 7.0.

\begin{figure}[t!]
    \centering
    \includegraphics[width=1.0\linewidth]{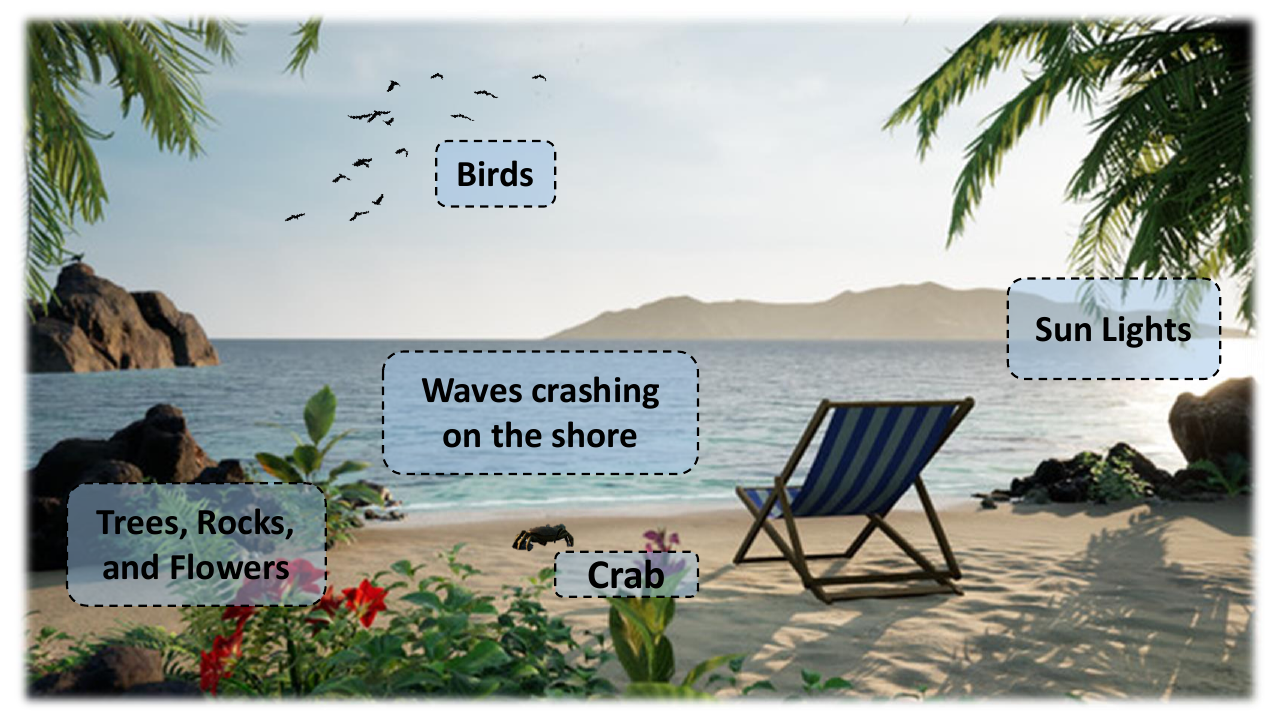}
    \caption{Meditation VR. This free application is available from Steam. We decide to use this application for our VR Meditation condition because it satisfies four out of five criteria: visual attractiveness, realism, user comfort, and navigation methods~\cite{zaharuddin2019factors}.}
    \label{Fig:VR_meditation_app}
\end{figure}

\paragraph{VR Meditation:} 
Participants in this condition experienced a virtual beach scene for 5 minutes in an immersive VR environment (see Fig.~\ref{Fig:VR_meditation_app}). 
For the scene, we used a free application from Steam, Meditation VR\footnote{https://store.steampowered.com/app/1301850/Meditation\_VR/}.
The scene had a sandy beach on an isolated island surrounded by trees and rocks. 
Birds flying high above the sea and small crabs passing between sand and rocks visually attracted the user.
Waves crashing on the shore and the wind blowing through the trees sounds were included to give the participants a sense of realism. 
However, interactions with these elements were not supported.
The participants were allowed to sit on a chair and take a break or move around in VR as shown in Fig.~\ref{fig:teaser}(b).
This condition satisfies four of the five criteria for better virtual meditation spaces suggested by Zaharuddin et al.~\cite{zaharuddin2019factors}: visual attractiveness, realism, user comfort, and navigation methods. A total of 13 participants (4 male, 9 female) experienced VR Meditation. Of these, 12 had prior VR experience, with an average self-reported VR familiarity score of 3.17.

\begin{figure*}[t!]
    \centering
    \includegraphics[width=1.0\linewidth]{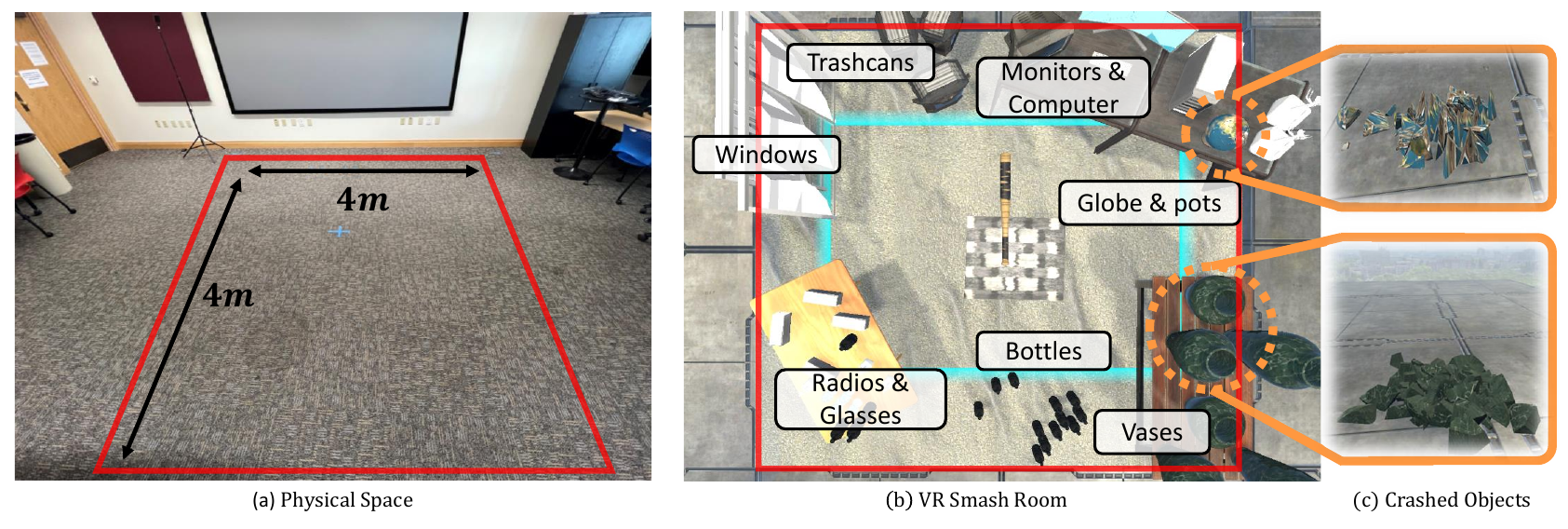}
    \caption{The VR tracking space was $4\,m\times4\,m$. This space was used for both VR treatment scenes. Participants could move around the VR scenes within this physical space. The VR Smash Room had various breakable objects including windows, vases, bottles, etc. }
    \label{Fig:Setting}
\end{figure*}

\paragraph{VR Smash Room:} 
This condition was designed to support the physical activity of the participants in a VR environment, specifically an object-breaking activity.
As illustrated in Fig.~\ref{Fig:Setting}, our VR Smash Room was a room-size virtual area with various 3D objects, such as windows, televisions, vases, desktop computers, and wine bottles.
The participants in the VR Smash Room were able to break the objects for 5 minutes by swinging a VR controller. 
The controller was visually represented as a baseball bat in the VR scene (Fig.~\ref{fig:teaser}(c)). 
Each 3D object had a different hitting velocity threshold ($3\,m/s$ or $6\,m/s$) to be broken when the participants swing the controller and strike it. 
For example, wine bottles and glasses, and windows could be broken with a swing faster than $3\,m/s$; and other objects (e.g., TV, computer, and vases) needed a swing faster than $6\,m/s$ to be destroyed. 
When an object was crashed, as shown in Fig~\ref{Fig:Setting}(c), a glass-breaking sound was played. An object having a higher hitting velocity threshold had a louder sound. Broken object pieces are created by using the cell fracture feature provided in Blender~\cite{Blender}. 14 participants (8 male, 6 female) experienced VR Smash Room. Among them, 12 participants had prior VR experience with an average VR familiarity score of 3.17.


\subsection{Hypotheses}
\label{Sec:Hypotheses}

Our study aims at investigating the effects of stress-management methods in VR, compared with the real world condition. The earlier studies showed that practicing VR meditation could be as effective as doing so in the real world to reduce stress levels. 
However, the effects of physical activity in VR (e.g., VR Smash Room) have not been investigated well.
While addressing our research questions in Section~\ref{sec_introduction}), we establish the following hypotheses for our study:

\begin{enumerate}[label=\textbf{H\arabic*.}]

\item The VR treatments (i.e., VR Smash Room and VR Meditation) will have a greater stress-relief effect than Sitting-in-Silence in the real environment because of the immersion and isolation of the VR experiences that resolve environmental distractions.

\item VR Smash Room will have a larger stress-relief effect compared to VR Meditation and Sitting-in-Silence because its physical activity provides users with a fun and anger expression experience.



\item The effects of stress-relief treatments will vary based on participants' personality traits, as susceptibility to psychological stressors differs among individuals~\cite{MONROE2016109}.

\end{enumerate}

\subsection{Apparatus and Study Space}
\label{Sec:Apparatus}

For the study, an immersive VR head-mounted display (HMD) was used---a Vive Pro Eye headset with a Vive wireless adapter and a Vive controller.
The headset has a viewing angle of $1440\times1600$ for each eye and a refresh rate of $90\,Hz$.
The wireless adapter offers near-zero latency\footnote{https://www.vive.com/us/accessory/wireless-adapter/}. 
The VR Smash Room scene was implemented in Unity 2020.3.29f1, and the project is publicly available via our GitHub repository\footnote{anonymized for review}.
The two virtual scenes (i.e., VR Meditation and VR Smash Room) ran on a Windows 10 desktop with an Intel Xeon W-2245 CPU ($3.90\,GHz$), $64\,GB$ of RAM and an Nvidia GeForce RTX 3090 graphics card.
The physical tracking space for the VR scenes was $4m\times4m$ as shown in Fig.~\ref{Fig:Setting}(a). Participants playing the VR Meditation or Smash Room were allowed to move around within this space.

\begin{figure}[t]
    \centering
    \includegraphics[width=.9\linewidth]{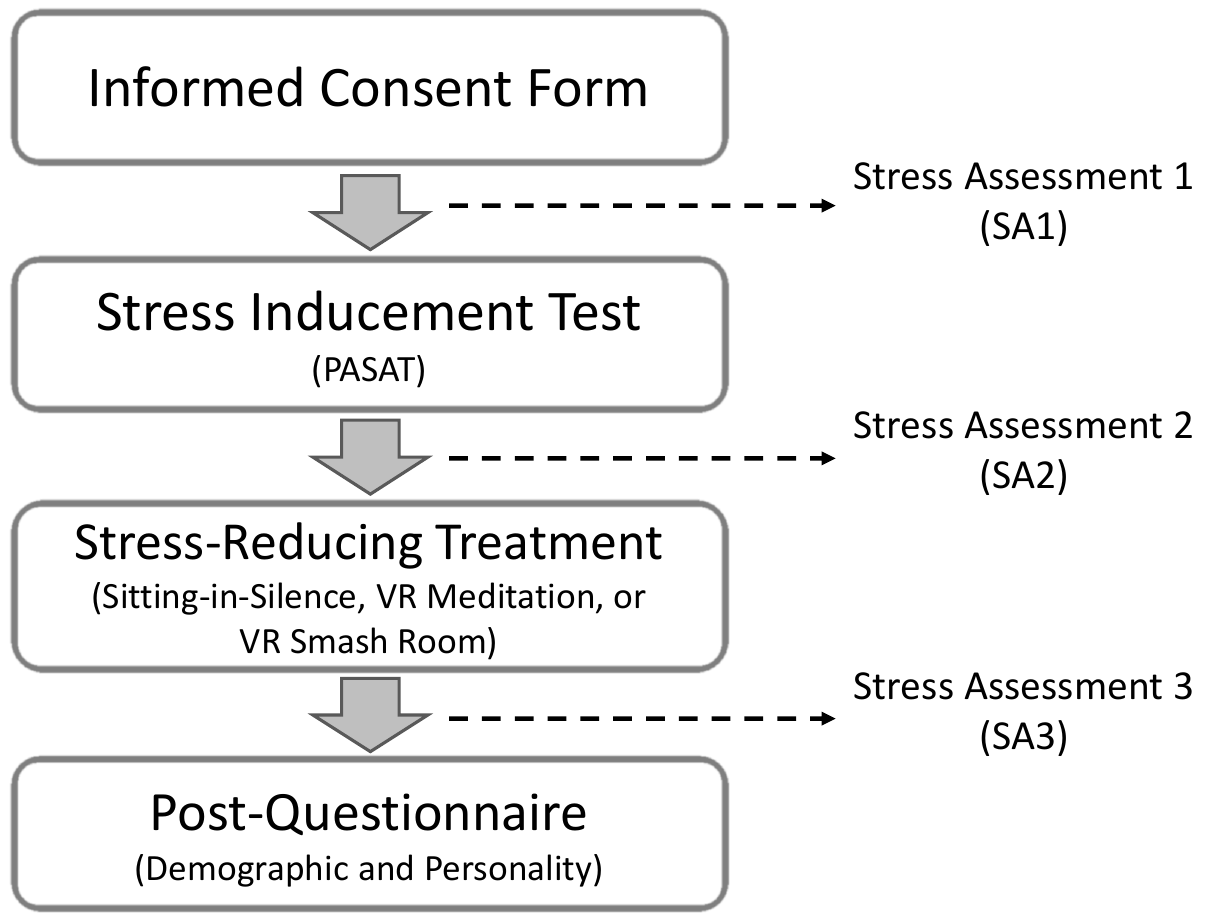}
    \caption{
    Study procedure: at the beginning of the study, participants were asked to read and agree on the informed consent form and evaluated their stress level by completing the STAI test. 
    Then they experienced two interventions: the stress-inducing task (PASAT) and the stress-reducing treatment. 
    After each intervention, the participants repeatedly reported their stress levels using the STAI test.
    }
    \label{Fig:Procedure}
\end{figure}

\subsection{Procedure}
\label{Sec:Procedure}

The study consisted of 1) reading and signing a consent form; 2) completing a series of stress-assessment questionnaires; 3) conducting a stress-inducing task; 4) experiencing a stress-reducing treatment; and 5) answering a post-questionnaire. 
Upon arrival at the study place, a participant was asked to read and sign the informed consent form to participate in the study according to the IRB protocol (IRB \#12971). 
The participant was first briefed on the study objectives and the procedure (Fig.~\ref{Fig:Procedure}). 
Then, the participant was asked to complete a stress assessment questionnaire, State-Trait Anxiety Inventory for Adults (STAI)~\cite{spielberger1983state}---the details about STAI are described in Section~\ref{Sec:Measures}.
The first STAI assessed the participant's initial stress level as a baseline.

Once the participant completed the first STAI, he/she was asked to perform a Paced Auditory Serial Addition Test (PASAT). PASAT is a cognitively challenging task, and it induces an increase in stress before experiencing our stress-reducing treatments.
PASAT is originally designed as a neuropsychological measure of cognitive processing speed, but it has been found that PASAT could be a good experimental tool to induce a sense of irritation, frustration, and anxiety for users~\cite{tombaugh2006comprehensive}.
The test consisted of four sets, and each set called out 60 numbers at regular intervals. While the 60 numbers were being called one by one, the participant must answer the sum of two numbers just heard and the previous number. The 60 numbers for the four sets were the same, but the interstimulus intervals decreased to 2.4s, 2.0s, 1.6s, and 1.2s as the sets progressed. PASAT took approximately 15 minutes. After PASAT, the participant was asked to complete the second STAI.

Next, the participant experienced one of the three stress-reducing treatments described in Section~\ref{Sec:StudyDesign} for 5 minutes and completed the third STAI.
When all the interventions were completed, the participant was asked to complete a post-questionnaire and the Big Five personality test~\cite{zillig2002we}. 
The post-questionnaire asked for demographic information, such as gender and age.
At last, the study instructor had a brief interview with the participant, asking about their general perception and feedback.

\subsection{Measures}
\label{Sec:Measures}

\paragraph{Ratio of Self-Reported Anxiety Scores}
Stress is typically caused by an external and existent source, and anxiety could be considered as a different type of stress that remains even if the source of stress is gone.
Among many different measuring tools, a self-reported measurement, STAI, is widely used to evaluate one's level of anxiety~\cite{spielberger1983state} STAI has two forms of questionnaires in a 4-point Likert scale: Y-1 and Y-2 forms. 
The Y-1 form evaluates the user's current/immediate feelings, while the Y-2 form is a set of statements to evaluate the user's general feelings.
In this study, we used the Y-1 form only to evaluate the participants' immediate anxiety levels before and after the interventions, i.e., PASAT and the stress relief treatments---see Stress Assessment (SA) 1, 2, and 3 in Fig.~\ref{Fig:Procedure}. The questionnaire had a total of twenty statements about anxiety perception, such as \textit{``I feel calm,''} \textit{``I am tense,''} and \textit{``I feel confused.''}\footnote{The STAI used in our study: \url{https://www.advancedassessments.co.uk/resources/Mental-Health-Test.pdf} (Accessed: 2023-03-22)}

The level of anxiety was decided by adding all item scores, which were rated on a 4-point Likert scale (e.g., from `Not at all' to `Very Much So'). 
As a result, the range of STAI scores is between 20 and 80. The higher scores the participant reports, the higher level of stress or anxiety they perceive: no or low anxiety (20--37), moderate anxiety (38--44), and high anxiety (45-80)~\cite{Kayikcioglu2017sat}.

STAI is designed to measure a respondent's anxiety objectively through self-reported short questions.
However, the evaluation of the perceived anxiety and its changes/fluctuations over time would still depend on the individual---e.g., someone might have more stress tolerance than another.
In this regard, we further processed the anxiety scores reported by the participants, to compare the relative stress-relief effects of our three treatments. We calculated ratios of STAI changes after the treatments by dividing them by the amount of the increased STAI score by the stress-inducing test (PASAT) phase, i.e., $(SA3-SA2)/(SA2-SA1)$. In this way, we considered individual participants' anxiety sensitivity. 

\paragraph{Personality Traits}
To identify potential relationships between participants' personalities and the stress-relief effects of the treatment interventions, their personalities were measured using the Big Five personality traits~\cite{zillig2002we}.
The Big Five personality traits model is a psychological model that explains human personality with five mutually independent factors.
The five factors are agreeableness, extraversion, conscientiousness, neuroticism, and openness.
Agreeableness is a behavioral characteristic about how one treats others, such as being cooperative and considerate.
Extraversion is a measurement that determines how actively people like to relate and interact with others.
Conscientiousness indicates a characteristic of being careful or diligent.
Neuroticism is a factor that determines how often an individual experiences negative emotions in daily life. 
Finally, openness represents how people judge new experiences. Among many Big Five test variations, the questionnaire used in our study had 50 statements like ``I am the life of the party'' and ``I am always prepared.''
The participants were asked to evaluate each statement on a 5-point Likert scale (1: very inaccurate to 5: very accurate). 
The statements had a positive or negative weight (e.g., $+1$ or $-1$) to evaluate personalities.

\paragraph{Movements and Smash Activities}
To identify potential relationships between a participant's activities and the stress-relieving effects, we measured how much a participant moved during the VR Smash Room experience.
We recorded the coordinates of the HMD and the VR controller in seconds. 
The HMD and VR controller movement distances were measured by calculating the Euclidean distances between their sequentially tracked coordinates in meters.
In addition, we recorded the number of smash hits and crashed virtual objects. 
The number of hits indicates how many times a participant swang the controller and hit the objects. 
The number of crashed virtual objects indicates how many virtual objects a participant successfully crashed by swinging the controller at a speed faster than a crash threshold.
Based on these two hit measurements, we calculated a critical hit score, which is a ratio of the number of crashed virtual objects over the number of hits.
\section{Results}
\label{Sec:Results}

To analyze the measurements collected during the experiment, we conducted multiple statistical analyses including mixed-subjects and between-subjects ANOVAs depending on the variable dependency, and linear regression analysis. 
The ANOVA tests were corrected with Bonferroni correction. The degrees of freedom were corrected using Greenhouse-Geisser correction to protect against violations of the sphericity assumption. 
Pairwise comparisons were conducted as post-hoc tests when ANOVAs showed significant effects. The significance level was at 5\% for all the tests.

\begin{figure}[t]
    \centering
    \includegraphics[width=.9\linewidth]{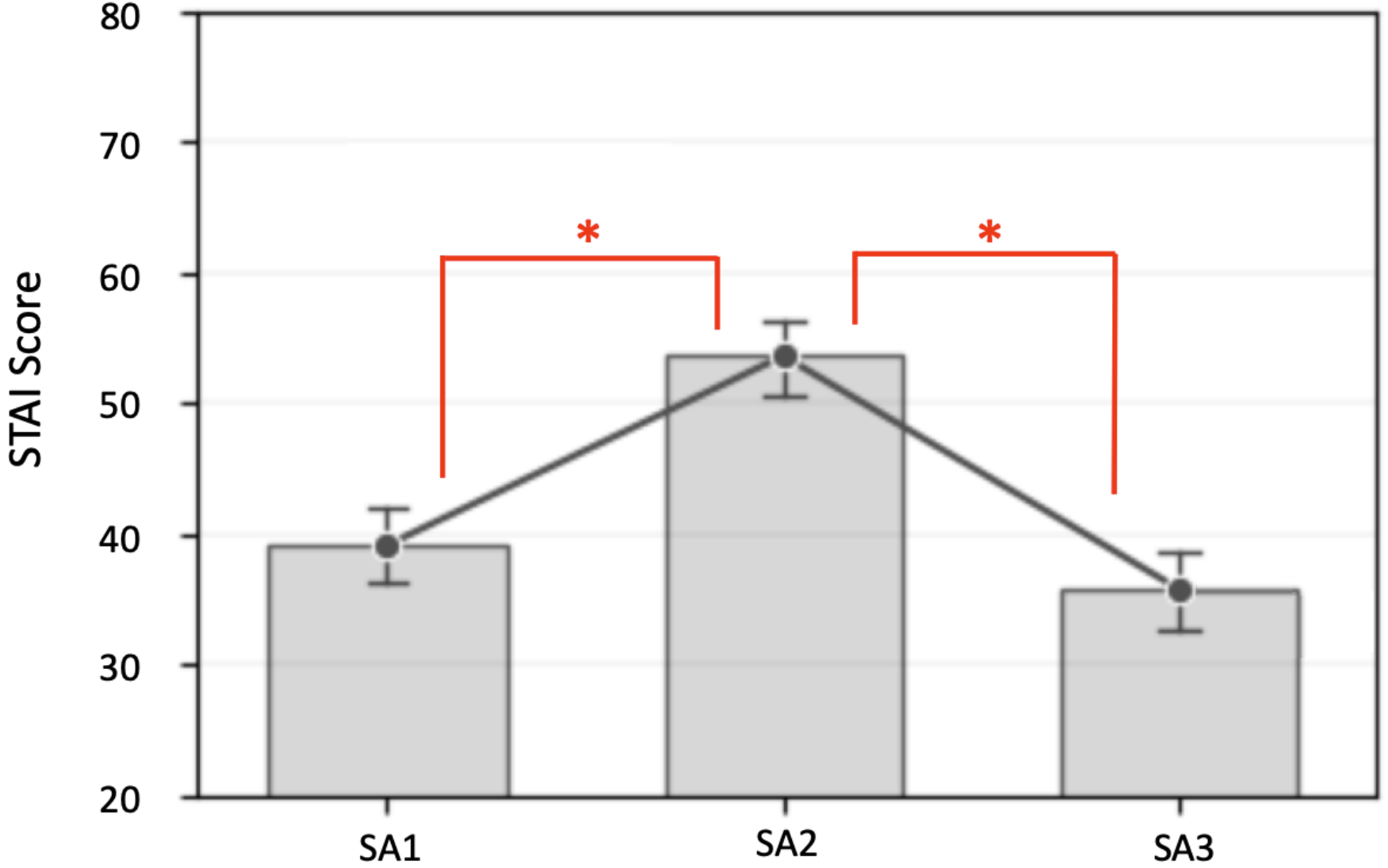}
    \caption{The results of the self-reported STAI anxiety scores along with means and 95\% CI error bars. This shows the STAI scores increased after PASAT, and decreased after the stress-reducing sessions in sequence. The range of the anxiety score is from 20 to 80. *($p<.05$).
    }
    \label{Fig:score_chart_by_stage}
\end{figure}

\begin{figure}[t]
    \centering
    \includegraphics[width=.95\linewidth]
    {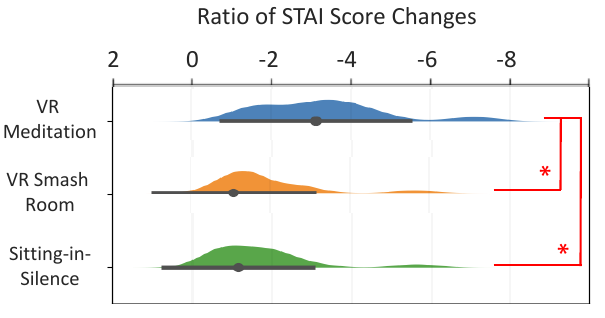}
    \caption{The stress-relief effects of different activities after the first treatment session. The averaged ratios of the anxiety score changes for the first treatment are shown along with 95\% CI error bars.
    The VR Meditation has a significantly larger drop in anxiety levels, compared to the VR Smash Room and the Sitting-in-Silence. *($p<.05$).}
    \label{Fig:changes_1stTreatment}
\end{figure}

\subsection{Self-Reported Anxiety Scores}
\label{Sec:Result_SelfReported}

We first analyzed the STAI scores to investigate how the anxiety scores changed over time throughout the interventions in the experiment, using a 3$\times$3 mixed-subjects (Interventions $\times$ Stress Assessments) ANOVA. 
An interaction effect ($p=.016$, $F(4, 74)=3.41$, ${\eta_p}^2=.155$) was reported. 
We found no simple effect of the interventions on the STAI scores. 
At SA1, there was no difference between the intervention groups ($p=.069$). 
Similarly, SA2 and SA3 showed no differences between the interventions, with $p=.728$ and $p=.143$ respectively. 
These results suggest that there are no differences in the STAI scores among the intervention groups.

Simple effects of the stress assessment on the STAI scores were disclosed in each intervention condition.
In the VR Meditation group, the STAI scores showed statistical differences over the assessments ($p=.001$, $F(2, 24)=38.6$, ${\eta_p}^2=.763$). 
SA2 ($M=55.1$, $SD=8.60$) showed a significantly higher score than SA1 ($M=43.1$, $SD=10.4$, $p<.001$) and SA3 ($M=31.8$, $SD=7.28$, $p<.001$).
A significant difference between SA1 and SA3 was also disclosed ($p=.005$).
The VR Smash group disclosed statistically different STAI scores over the assessments ($p=.001$, $F(2, 26)=31.6$, ${\eta_p}^2=.708$). 
SA2 ($M=53.5$, $SD=9.63$) had a higher score than SA1 ($M=34.6$, $SD=10.8$, $p<.001$) and SA3 ($M=35.3$, $SD=7.86$, $p<.001$).
However, no difference between SA1 and SA3 was reported ($p=1.00$).
Similarly, the participants in the Sitting-in-Silence group showed statistically different STAI scores over the assessments ($p=.001$, $F(2, 24)=22.3$, ${\eta_p}^2=.650$). 
SA2 ($M=56.9$, $SD=8.45$) was significantly higher than SA1 ($M=37.2$, $SD=6.82$, $p<.001$) and SA3 ($M=39.6$, $SD=12.6$, $p<.001$), but there was no difference between SA1 and SA3 ($p=1.000$).

The results revealed a main effect on the STAI scores (${{p}<.001}$, $F(2, 74)=82.2$, ${\eta_p}^2=.690$), which means that there is a statistical difference in the reported anxiety scores over the assessments.
It is shown in Fig.~\ref{Fig:score_chart_by_stage}.
Its pairwise comparisons revealed that SA2 ($M=55.1$, $SD=8.81$) was significantly higher than SA1 ($M=38.3$, $SD=9.95$; ${{p}<.001}$). 
This implies that PASAT could induce a significant increase in the participant's anxiety level, and we only included those with increased stress levels.
The comparisons also showed that SA3 ($M=35.6$, $SD=9.80$) was significantly lower than SA2 (${{p}<.001}$). 
We found no difference between SA1 and SA3 (${{p}=.302}$). 
We found no main effect on the interventions (${{p}=.460}$).


\begin{figure*}[ht]
    \centering
    \includegraphics[width=1\linewidth]{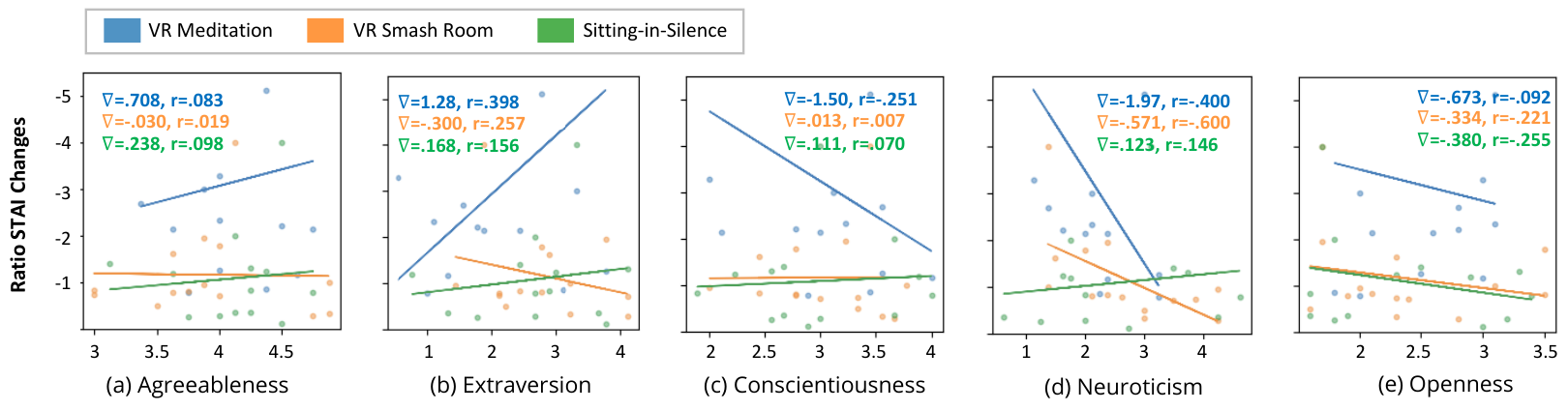}
    \caption{The linear regression results between the personality traits and the ratio of STAI changes after the first interventions are presented. The slopes ($\Delta$) and correlation strengths (r) are presented (r-value $\ge$ 0.6: strong, $\ge$ 0.4: moderate, $\ge$ 0.2: weak, $<$ 0.2: very weak). }
    \label{Fig:normalized_changes_by_personalities}
    \vspace{-.4cm}
\end{figure*}

\begin{figure}[t]
    \centering
    \includegraphics[width=1\linewidth]{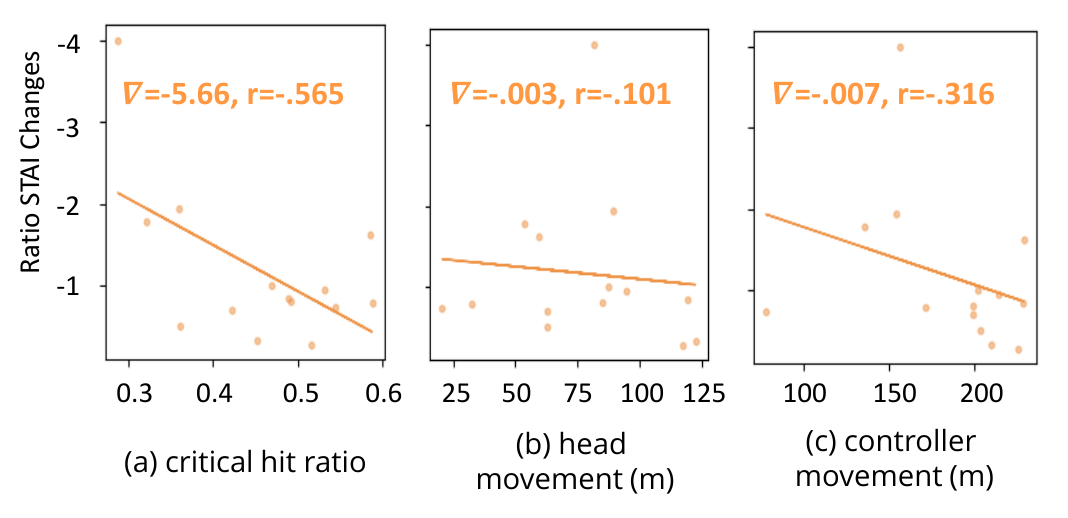}
    \caption{Ratio STAI Changes by user activities in VR Smash Room. The activities have negative relationships to the stress relief effects.}
    \label{Fig:normalized_changes_by_activity}
    \vspace{-.4cm}
\end{figure}

Next, we report the effects of the stress-reducing treatments on the ratios of the STAI score changes between SA2 and SA3, i.e., $(SA3-SA2)/(SA2-SA1)$, using a between-subjects ANOVA. 
Fig.~\ref{Fig:changes_1stTreatment} shows the ratios of STAI score changes along with their 95\% CI.
The results showed a significant difference between the treatments ($p=.028$, $F(2, 37)=3.95$, ${\eta_p}^2=.176$).
In the pairwise comparisons, VR Meditation ($M=-3.16, SD=1.22$) had a significantly larger stress relief effect than VR Smash Room ($M=-1.17, SD=.957, {{p}=.021}$) and Sitting-in-Silence ($M=-1.09, SD=1.03, {{p}=.018}$).
However, we found no significant difference between VR Smash Room and Sitting-in-Silence ($p=.926$).


\subsection{Personality Traits}
\label{Sec:Results_Personality}

To examine potential relationships between the participants' personality traits and the effects of the stress-reducing treatments, we conducted linear regression analyses using the ratios of STAI score changes after the treatment. 
The measured personality trait scores were used to fit the linear regression models with the ratios of the STAI changes. 
The estimated slope coefficients ($\nabla$) and Pearson correlation coefficient ($r$) are reported together. 
The details of the linear regression results are shown in Fig.~\ref{Fig:normalized_changes_by_personalities}.

First of all, we found a substantial negative relationship between neuroticism and VR Smash Room ($\nabla=-.571$, $r=-.600$).
This indicates that having the VR Smash Room experience is less effective in reducing stress for the participants who tend to experience negative emotions more easily, but more effective for those who are more emotionally stable and less reactive to stress.
The VR Smash Room also had a negative association with extraversion ($\nabla=-.300$, $r=-.257$) and openness ($\nabla=-.330$, $r=-.221$), but the correlations were weak.

Regarding VR Meditation, moderate relationships to the traits are shown. 
It has a positive relationship with extraversion ($\nabla=1.28$, $r=.398$) and a negative relationship with neuroticism ($\nabla=-1.97$, $r=-.400$). 
In other words, VR Meditation is more effective for extravert and less negative people. 
A weak negative association between conscientiousness and VR Meditation is also found.

Sitting-in-Silence only has weak relationships with extraversion ($\nabla=.69$, $r=.156$), neuroticism ($\nabla=.123$, $r=.146$), and openness ($\nabla=-.380$, $r=-.255$).
Other than these, we found only very weak relationships between the treatment effects and the traits.

\subsection{Movements and Smash Activities}
The relationships between participants' activities and the stress-relief effects are examined, specifically for the VR Smash Room treatment. 
We conducted linear regression to examine the relationships and the results are shown in Fig.~\ref{Fig:normalized_changes_by_activity}.
The stress-relief effects have a moderate negative association with the critical hit ratio ($\nabla=-5.66$, $r=-.565$). 
This shows that the stress-relief effect of the VR Smash Room is stronger for the participants who broke fewer virtual objects than those who broke more objects.
There are also weak and very weak negative relationships with the controller movement ($\nabla=-.007$, $r=-.316$) and the head movement ($\nabla=-.003$, $r=-.101$).
\section{Discussion}
\label{Sec:Discussion}

The results of our formal user study show that the stress levels changed over the interventions.
This means that our stress-inducing PASAT task and three stress-reducing treatments influenced the participant-perceived stress level.
In this section, we summarize and justify the details of our findings based on the results, and discuss the implications of using different VR activities for stress relief.

\begin{figure*}[t!]
    \centering
    \includegraphics[width=1\linewidth]{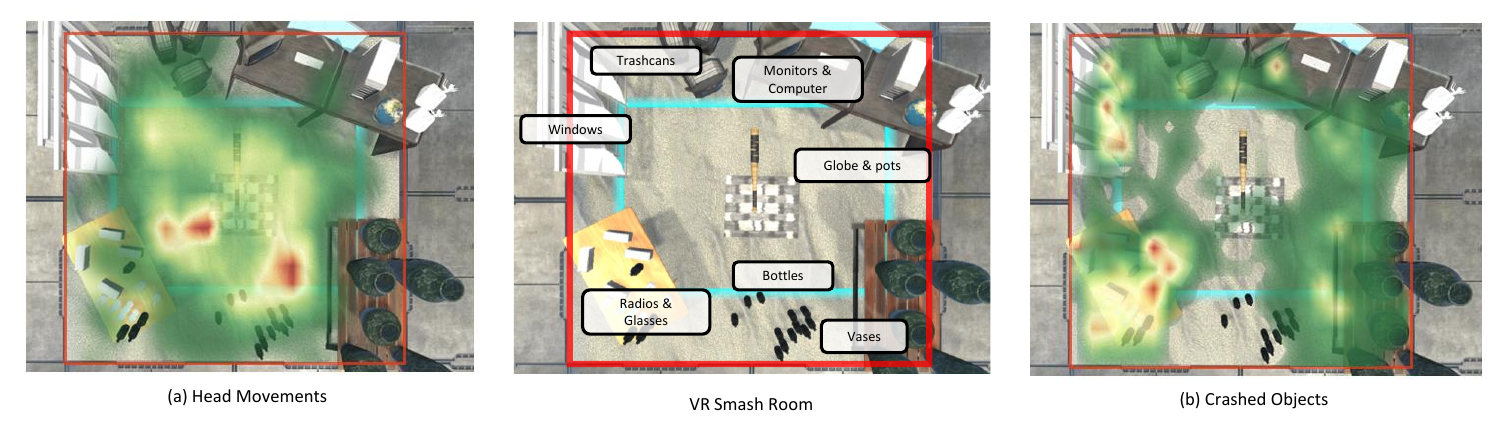}
    \caption{Participants' activities in VR Smash Room. (a) the heatmap reveals how the participants move around the virtual space. Red indicates the participants stayed longer time. (b) the heatmap shows the participants broke more objects which have a low-velocity threshold and a soft breaking sound.}
    \label{Fig:smashRoom-activity}
    \vspace{-.4cm}
\end{figure*}

\paragraph{\textbf{VR Meditation is more effective for stress relief than VR Smash Room and Sitting-in-Silence.}}

Given the findings from previous literature described in Section~\ref{Sec:RelatedWork}, we established two hypotheses (\textbf{H1} and \textbf{H2}) related to the effects of three stress-reducing activities. In H1, we expected both virtual activities (i.e., VR Meditation and VR Smash Room) would have a stronger effect on reducing stress compared to Sitting-in-Silence because the immersive virtual environment could make the participants isolated and detached from the stressor in reality. 
In addition, we hypothesized that VR Smash Room would be more effective than VR Meditation due to the active and dynamic physical movement in H2. Our results revealed that VR Meditation was more effective to reduce the level of stress in terms of the ratios of the STAI score changes than VR Smash Room and Sitting-in-Silence with a statistical significance. Furthermore, only SA3 after VR Meditation is significantly lower than SA1. These findings partially support our H1, but do not support H2. 
Although the findings are different from what we expected, we could understand and justify them in other clinical and psychological aspects, as well as based on the literature that addresses the lack of consistency in the effects of destructive activities.

The after-study interview results help us understand deeper about the interventions. First of all, in our study, VR Meditation provided a natural environment through an immersive VR device.
The participants commented that such a natural environment (e.g., the immersive beach scene) allowed them to feel relaxed and easy to calm their minds.
They also reported that exploring the wildlife around the beach with the relaxing sounds and watching little crabs and lizards were fun. 
This is also aligned with the findings in the previous studies that immersion and presence in a natural environment via VR have a mental restorative effect to resolve environmental distractions and divert their attention from the real world~\cite{iyendo2016therapeutic, zaharuddin2019factors}.

On the other hand, a few participants reported that VR Smash Room was entertaining and reduced stress, but it started to get stale soon and unpleasant. 
One of them mentioned that she found it difficult to relax in the VR Smash Room because she felt that breaking objects in the VR Smash Room was perceived as a task that she needed to perform.
Several participants who played VR Smash Room also commented that hearing the sound of objects crashing was harsh to their ears.
Here we share some participants' comments related to the negative aspect of the VR Smash Room.
\begin{quote}
\textit{P15: ``I found the sounds in the (VR) Smash Room to be very unsatisfying and loud, so they caused me to have a headache after the VR session was over.''}

\textit{P19: ``The (VR) smash room was fun, but it started to get old. I started to look for new things to do.''}

\textit{P32: ``It (VR Smash Room) helped me feel a little less tense, but I still felt a little stressed.''}

\end{quote}
Based on the comments and our observation, expressing anger by breaking objects in VR Smash Room may not be the best way to relieve stress, but possibly stimulate the participant's sensory perception with unpleasant sound and visual outcomes. 
Such sensory stimuli could apparently lead to different types of cognitive stress, and possibly make the users angrier and more aggressive~\cite{xie2009clinical, alvarsson2010stress}. 

Lastly, the type of stress inducer may play a role in the effects of different stress-reducing activities. 
In our study, we used a cognitive task, PASAT, to increase the participants' stress levels.
This type of cognitive stress could be more effectively managed by mental activities that moderate the cognitive state like VR Meditation, whereas the extensive physical activities in VR Smash Room may not be appropriate~\cite{goyal2014meditation, hasan2013violent}. In general, our results support the therapeutic potential of VR Meditation for anxiety management and stress reduction programs among the three treatments in the study.

\paragraph{\textbf{A brief 5-minute treatment experience can reduce stress significantly.}}

Our results revealed that having a brief 5-minute treatment session was effective enough for the participants to recover from short-term induced stress.
In our study, the participants initially had low-moderate anxiety scores (SA1: $M=38.2$) as a baseline before the PASAT, and the anxiety increased significantly to the level of high anxiety after the PASAT (SA2: $M=55.1$).
After that, the 5-minute treatment significantly reduced the participants' stress (SA3: $M=35.6$) similar to the prior stress level (SA1) or even slightly lower. Especially, the effect of VR Meditation was conspicuous in reducing stress.
Although there could be some individual differences, the literature addressed that 10--20 minutes per day would be a reasonable amount of time to meditate for stress relief~\cite{Hart2007}.
In this sense, our study also showed that VR Meditation could be a viable method to manage daily stress even within a shorter time.



\paragraph{\textbf{Personality traits are associated with the stress-relief effects.}}
Stress is a sense of emotional or physical tension caused by external sources, but stress perception and resilience can vary significantly from person to person~\cite{Southwick2016}. It is known that personality characteristics could explain susceptibility to stress and vulnerability~\cite{digangi2013pretrauma, tosevski2010personality}. Thus, we hypothesized that there would be correlations between different personality traits and the effects of three treatments on stress relief (\textbf{H3}). Our results revealed various substantial and moderate correlations between the participants' personality traits and the stress-relief effects supporting H3.

More specifically, the stress-relief effect has a positive association with the extraversion trait and a negative association with the neuroticism trait in VR Meditation.
As shown in Fig.~\ref{Fig:normalized_changes_by_personalities}, the slopes of extraversion and neuroticism are larger than 1 with a moderate correlation.
This implies that VR Meditation was more effective for the participants who exhibited a higher extraversion trait and a lower level of neuroticism.
On the other hand, the effects of VR Smash Room were negatively associated with the neuroticism trait.
This aligns with previous findings that showed a correlation of neuroticism factor with stress-related psychopathologies~\cite{barlow2014origins, brown2011direct, kendler2010genetic}.
This could be explained by some participants' comments about negative perceptions of the high-pitched glass-breaking sounds and the repeated same smashing activity in VR Smash Room, which were described as irritating and discomfort.
In other words, participants with a high neuroticism score could be influenced by this negative perception more easily.
Unlike the VR treatments, Sitting-in-Silence has no significant correlations with personality traits.
In summary, our findings suggest that VR-based stress treatment applications or interventions should be designed while taking the user's personal profile (e.g., personality) into account.

\paragraph{\textbf{The more activities in VR Smash Room, the less stress relief effect.}}

Although we did not have specific hypotheses regarding participants' physical activities, we analyzed the correlations between the amounts of activities and the stress relief effects in VR Smash Room to understand what other factors could negatively affect the stress relief effect. 
Interestingly, our results showed that active breaking activities in VR Smash Room did not help reduce the participants' stress.
This indicates that engaging in a destructive activity like VR Smash Room with more physical movements may have less of a relaxing effect.
It is consistent with the participants' negative comments about VR Smash Room, which we covered above (P15, P19, and P32).

For more details, the participants' activities in VR Smash Room were calculated by their head position movements and the number of their crashed objects. 
Heatmap representations for the head movement and the number of crashed objects are shown in Fig.~\ref{Fig:smashRoom-activity}.
Fig.~\ref{Fig:smashRoom-activity}(a) shows a heatmap of the participants' movement in the VR Smash Room.
The heatmap highlights two highest frequency spots: one is in front of the light brown desk, which is roughly the center of the smash room, and another is in front of the vases. 
We observed that the participants successfully broke many objects (e.g., wine bottles and glasses, and radios) on the table but they did not break many vases.
Fig.~\ref{Fig:smashRoom-activity}(b) shows a heatmap of the crashed objects by the participants. 
We could observe that the participants broke window objects more than other objects even though they did not stay in the nearby position for long.
The velocity threshold of the objects like vases, TV, or computers to be broken was double of other objects, such as windows, wine bottles and glasses, and radios.
Given these different thresholds and the heatmap result, we infer that the participants had a tendency to be located near objects that can be easily broken.
There could be two interpretations for this tendency.
Firstly, the participants might have wanted to avoid crashing objects that cause a loud crash sound.
We consider that the stress relief effect of VR Smash Room is adversely affected by the objects that are hardly broken and large harshed sounds. 
Secondly, the experience in which they failed to break objects could interrupt the stress relief effect and would increase their stress level conversely.

\subsection{Limitations and Future Work}
\label{Sec:Limitations}
We covered interesting findings of our study and discussed the implications of the findings, but we also admit there could be some potential limitations, which open up valuable future research directions.
This section discusses such limitations and future work. 

One of the limitations may stem from our stress measurement approach. In our work, the stress levels were measured by using the self-evaluation method (STAI) after the interventions, such as PASAT and the treatments. While STAI is a practical, valid, and widely-used method to measure one’s stress level, it is still highly subjective. 
We used it three times, similar to other earlier studies that used STAI repeatedly in their research.
However, repeating this kind of subjective questionnaire multiple times may have an unexpected impact on the results.
In addition, STAI does not allow us to monitor stress level changes in real-time while a participant is experiencing a treatment. These limitations could be overcome through different measurements, such as an electrocardiogram and skin conductance. 

Unbalanced population distribution is the second limitation. In our study, the gender distribution within each condition is not fully balanced, with one condition having more male participants than female participants and the others having more female participants than male participants with a limited age range between 18 and 22.
In order to examine the general effects of VR on stress relief, it is necessary to conduct follow-up research with various demographic groups, such as ages, genders, races, and educational backgrounds, with a balanced population distribution for the treatment conditions. 

Finally, the effects of long-term VR treatments are still unknown. More specifically, our study evaluated the short-term effects of the VR treatments compared to the traditional method, Sitting-in-Silence, and clearly demonstrated that the VR treatments helped participants manage stress. However, to investigate the long-term effects of those treatments, follow-up studies are required.

Future research would examine how effective different VR experiences are in relation to various types of stressors. 
In the real world, unlike our study, various types of stressors exist including social, organizational, and environmental stressors. Various VR games and experiences are available as potential methods for stress relief. Future research should examine the effectiveness of various VR activities like flying in the sky, underwater exploration, and playing with virtual animals as well as the effectiveness of human senses in VR (e.g., haptic and olfactory feedback) on stress reduction.
\section{Conclusion}
\label{Sec:Conclusion}

In this paper, we conducted a formal user study with 40 participants to evaluate the three treatments (VR Meditation, VR Smash Room, and Sitting-in-Silence) on stress relief.  
The participants completed two interventions including PASAT and the stress treatment. 
The results of our study confirmed that all three treatments could reduce their stress levels, but VR Meditation was the most effective method. 
We discussed the implications of our findings while covering not only the treatment type, but also the association with personality traits, and the degree of physical activity.
Our results also revealed that participants could have some negative perceptions of VR Smash Room due to the destructive nature of the activity and unpleasant sound effects during the object-breaking activity, which possibly impact the stress relief effect.

\acknowledgments{
We acknowledge that this research is partially supported by the Natural Sciences and Engineering Research Council of Canada (NSERC), [RGPIN-2022-03294].
}

\bibliographystyle{abbrv-doi}

\bibliography{0_main}

\begin{thebibliography}{10}

\bibitem{aiken2015posttraumatic}
M.~P. Aiken and M.~J. Berry.
\newblock Posttraumatic stress disorder: possibilities for olfaction and
  virtual reality exposure therapy.
\newblock {\em Virtual Reality}, 19(2):95--109, 2015.

\bibitem{alvarsson2010stress}
J.~J. Alvarsson, S.~Wiens, and M.~E. Nilsson.
\newblock Stress recovery during exposure to nature sound and environmental
  noise.
\newblock {\em International journal of environmental research and public
  health}, 7(3):1036--1046, 2010.

\bibitem{baer2008construct}
R.~A. Baer, G.~T. Smith, E.~Lykins, D.~Button, J.~Krietemeyer, S.~Sauer,
  E.~Walsh, D.~Duggan, and J.~M.~G. Williams.
\newblock Construct validity of the five facet mindfulness questionnaire in
  meditating and nonmeditating samples.
\newblock {\em Assessment}, 15(3):329--342, 2008.

\bibitem{barlow2014origins}
D.~H. Barlow, K.~K. Ellard, S.~Sauer-Zavala, J.~R. Bullis, and J.~R. Carl.
\newblock The origins of neuroticism.
\newblock {\em Perspectives on Psychological Science}, 9(5):481--496, 2014.

\bibitem{Bennett2017}
K.~Bennett.
\newblock Rage rooms not a good idea: Are you “letting off steam” in a rage
  room or motivating future aggression?
\newblock {\em Psychology Today}, pp. 1--2, Mar. 2017.

\bibitem{berto2014role}
R.~Berto.
\newblock The role of nature in coping with psycho-physiological stress: a
  literature review on restorativeness.
\newblock {\em Behavioral sciences}, 4(4):394--409, 2014.

\bibitem{Brown2012}
G.~W. Brown and T.~Harris.
\newblock {\em Social origins of depression: A study of psychiatric disorder in
  women}.
\newblock Routledge, 2012.

\bibitem{brown2011direct}
T.~A. Brown and A.~J. Rosellini.
\newblock The direct and interactive effects of neuroticism and life stress on
  the severity and longitudinal course of depressive symptoms.
\newblock {\em Journal of Abnormal Psychology}, 120(4):844, 2011.

\bibitem{campos2016meditation}
D.~Campos, A.~Cebolla, S.~Quero, J.~Bret{\'o}n-L{\'o}pez, C.~Botella, J.~Soler,
  J.~Garc{\'\i}a-Campayo, M.~Demarzo, and R.~M. Ba{\~n}os.
\newblock Meditation and happiness: Mindfulness and self-compassion may mediate
  the meditation--happiness relationship.
\newblock {\em Personality and individual differences}, 93:80--85, 2016.

\bibitem{chambers2008impact}
R.~Chambers, B.~C.~Y. Lo, and N.~B. Allen.
\newblock The impact of intensive mindfulness training on attentional control,
  cognitive style, and affect.
\newblock {\em Cognitive therapy and research}, 32(3):303--322, 2008.

\bibitem{chandrasiri2020virtual}
A.~Chandrasiri, J.~Collett, E.~Fassbender, and A.~De~Foe.
\newblock A virtual reality approach to mindfulness skills training.
\newblock {\em Virtual Reality}, 24:143--149, 2020.

\bibitem{Cohen2007}
S.~Cohen, D.~Janicki-Deverts, and G.~E. Miller.
\newblock Psychological {Stress} and {Disease}.
\newblock {\em JAMA}, 298(14):1685, Oct. 2007. doi: {{%
10\hspace{.1pt}\discretionary{.}{%
}{.}\hspace{.4pt}1001\discretionary{/}{%
}{/}jama\hspace{.1pt}\discretionary{.}{%
}{.}\hspace{.4pt}298\hspace{.1pt}\discretionary{.}{%
}{.}\hspace{.4pt}14\hspace{.1pt}\discretionary{.}{%
}{.}\hspace{.4pt}1685}}


\bibitem{difede2014d}
J.~Difede, J.~Cukor, K.~Wyka, M.~Olden, H.~Hoffman, F.~S. Lee, and M.~Altemus.
\newblock D-cycloserine augmentation of exposure therapy for post-traumatic
  stress disorder: a pilot randomized clinical trial.
\newblock {\em Neuropsychopharmacology}, 39(5):1052--1058, 2014.

\bibitem{digangi2013pretrauma}
J.~A. DiGangi, D.~Gomez, L.~Mendoza, L.~A. Jason, C.~B. Keys, and K.~C. Koenen.
\newblock Pretrauma risk factors for posttraumatic stress disorder: A
  systematic review of the literature.
\newblock {\em Clinical Psychology Review}, 33(6):728--744, 2013.

\bibitem{dooris2013expert}
M.~Dooris.
\newblock Expert voices for change: bridging the silos—towards healthy and
  sustainable settings for the 21st century.
\newblock {\em Health \& Place}, 20:39--50, 2013.

\bibitem{Dorsey2022}
A.~Dorsey, E.~Scherer, R.~Eckhoff, and R.~Furberg.
\newblock Measurement of human stress: A multidimensional approach.
\newblock {\em RTI Press}, 2022.

\bibitem{Filaire1996}
E.~Filaire, P.~Duche, G.~Lac, and A.~Robert.
\newblock Saliva cortisol, physical exercise and training: influences of
  swimming and handball on cortisol concentrations in women.
\newblock {\em European journal of applied physiology and occupational
  physiology}, 74(3):274--278, 1996.

\bibitem{flett2019mobile}
J.~A. Flett, H.~Hayne, B.~C. Riordan, L.~M. Thompson, and T.~S. Conner.
\newblock Mobile mindfulness meditation: a randomised controlled trial of the
  effect of two popular apps on mental health.
\newblock {\em Mindfulness}, 10(5):863--876, 2019.

\bibitem{freeman2018automated}
D.~Freeman, P.~Haselton, J.~Freeman, B.~Spanlang, S.~Kishore, E.~Albery,
  M.~Denne, P.~Brown, M.~Slater, and A.~Nickless.
\newblock Automated psychological therapy using immersive virtual reality for
  treatment of fear of heights: a single-blind, parallel-group, randomised
  controlled trial.
\newblock {\em The Lancet Psychiatry}, 5(8):625--632, 2018.

\bibitem{gebara2015virtual}
C.~M. Gebara, T.~P.~d. Barros-Neto, L.~Gertsenchtein, and F.~Lotufo-Neto.
\newblock Virtual reality exposure using three-dimensional images for the
  treatment of social phobia.
\newblock {\em Brazilian Journal of Psychiatry}, 38:24--29, 2015.

\bibitem{goyal2014meditation}
M.~Goyal, S.~Singh, E.~M. Sibinga, N.~F. Gould, A.~Rowland-Seymour, R.~Sharma,
  Z.~Berger, D.~Sleicher, D.~D. Maron, H.~M. Shihab, et~al.
\newblock Meditation programs for psychological stress and well-being: a
  systematic review and meta-analysis.
\newblock {\em JAMA internal medicine}, 174(3):357--368, 2014.

\bibitem{grossman2004mindfulness}
P.~Grossman, L.~Niemann, S.~Schmidt, and H.~Walach.
\newblock Mindfulness-based stress reduction and health benefits: A
  meta-analysis.
\newblock {\em Journal of psychosomatic research}, 57(1):35--43, 2004.

\bibitem{Hart2007}
J.~Hart.
\newblock {Clinical Applications for Meditation: A Review and Recommendations}.
\newblock {\em Alternative and Complementary Therapies}, 13(1):24--29, 2007.
  doi: {{%
10\hspace{.1pt}\discretionary{.}{%
}{.}\hspace{.4pt}1089\discretionary{/}{%
}{/}act\hspace{.1pt}\discretionary{.}{%
}{.}\hspace{.4pt}2006\hspace{.1pt}\discretionary{.}{%
}{.}\hspace{.4pt}13104}}


\bibitem{hasan2013violent}
Y.~Hasan, L.~B{\`e}gue, and B.~J. Bushman.
\newblock Violent video games stress people out and make them more aggressive.
\newblock {\em Aggressive behavior}, 39(1):64--70, 2013.

\bibitem{Herbert2020}
C.~Herbert, F.~Meixner, C.~Wiebking, and V.~Gilg.
\newblock Regular physical activity, short-term exercise, mental health, and
  well-being among university students: the results of an online and a
  laboratory study.
\newblock {\em Frontiers in psychology}, 11:509, 2020.

\bibitem{irving2009cultivating}
J.~A. Irving, P.~L. Dobkin, and J.~Park.
\newblock Cultivating mindfulness in health care professionals: A review of
  empirical studies of mindfulness-based stress reduction (mbsr).
\newblock {\em Complementary therapies in clinical practice}, 15(2):61--66,
  2009.

\bibitem{iyendo2016therapeutic}
T.~O. Iyendo, P.~C. Uwajeh, and E.~S. Ikenna.
\newblock The therapeutic impacts of environmental design interventions on
  wellness in clinical settings: a narrative review.
\newblock {\em Complementary therapies in clinical practice}, 24:174--188,
  2016.

\bibitem{jha2010examining}
A.~P. Jha, E.~A. Stanley, A.~Kiyonaga, L.~Wong, and L.~Gelfand.
\newblock Examining the protective effects of mindfulness training on working
  memory capacity and affective experience.
\newblock {\em Emotion}, 10(1):54, 2010.

\bibitem{jones2017real}
M.~Jones, A.~Taylor, Y.~Liao, S.~S. Intille, and G.~F. Dunton.
\newblock Real-time subjective assessment of psychological stress: associations
  with objectively-measured physical activity levels.
\newblock {\em Psychology of sport and exercise}, 31:79--87, 2017.

\bibitem{kaplan2021impact}
R.~Kaplan-Rakowski, K.~R. Johnson, and T.~Wojdynski.
\newblock The impact of virtual reality meditation on college students’ exam
  performance.
\newblock {\em Smart Learning Environments}, 8(1):1--15, 2021.

\bibitem{Kayikcioglu2017sat}
O.~Kayikcioglu, S.~Bilgin, G.~Seymenoglu, and A.~Deveci.
\newblock {State and Trait Anxiety Scores of Patients Receiving Intravitreal
  Injections}.
\newblock {\em Biomedicine Hub}, 2(2):1--5, 2017. doi: {{%
10\hspace{.1pt}\discretionary{.}{%
}{.}\hspace{.4pt}1159\discretionary{/}{%
}{/}000478993}}


\bibitem{kendler2010genetic}
K.~S. Kendler and J.~Myers.
\newblock The genetic and environmental relationship between major depression
  and the five-factor model of personality.
\newblock {\em Psychological medicine}, 40(5):801--806, 2010.

\bibitem{kim2021effect}
H.~Kim, D.~J. Kim, S.~Kim, W.~H. Chung, K.-A. Park, J.~D. Kim, D.~Kim, M.~J.
  Kim, K.~Kim, and H.~J. Jeon.
\newblock Effect of virtual reality on stress reduction and change of
  physiological parameters including heart rate variability in people with high
  stress: an open randomized crossover trial.
\newblock {\em Frontiers in psychiatry}, 12, 2021.

\bibitem{Lehrer1994}
P.~M. Lehrer, R.~Carr, D.~Sargunaraj, and R.~L. Woolfolk.
\newblock {Stress management techniques: Are they all equivalent, or do they
  have specific effects?}
\newblock {\em Biofeedback and Self-Regulation}, 19(4):353--401, dec 1994. doi:
  {{%
10\hspace{.1pt}\discretionary{.}{%
}{.}\hspace{.4pt}1007\discretionary{/}{%
}{/}BF01776735}}


\bibitem{li2020effects}
C.~Li, C.~Sun, M.~Sun, Y.~Yuan, and P.~Li.
\newblock Effects of brightness levels on stress recovery when viewing a
  virtual reality forest with simulated natural light.
\newblock {\em Urban Forestry \& Urban Greening}, 56:126865, 2020.

\bibitem{linehan2014dbt}
M.~Linehan.
\newblock {\em DBT? Skills training manual}.
\newblock Guilford Publications, 2014.

\bibitem{Lovibond1993}
S.~Lovibond and P.~Lovibond.
\newblock Manual for the depression anxiety stress scales (dass). psychology
  foundation monograph.
\newblock {\em Sydney, Australia: University of New South Wales}, 1993.

\bibitem{mantovani2003virtual}
F.~Mantovani, G.~Castelnuovo, A.~Gaggioli, and G.~Riva.
\newblock Virtual reality training for health-care professionals.
\newblock {\em CyberPsychology \& Behavior}, 6(4):389--395, 2003.

\bibitem{Mariotti2015}
A.~Mariotti.
\newblock The effects of chronic stress on health: new insights into the
  molecular mechanisms of brain-body communication.
\newblock {\em Future Science OA}, 1(3):fso.15.21, Nov. 2015. doi: {{%
10\hspace{.1pt}\discretionary{.}{%
}{.}\hspace{.4pt}4155\discretionary{/}{%
}{/}fso\hspace{.1pt}\discretionary{.}{%
}{.}\hspace{.4pt}15\hspace{.1pt}\discretionary{.}{%
}{.}\hspace{.4pt}21}}


\bibitem{Martin2016}
C.~Martin.
\newblock Anger rooms: A smashing new way to relieve stress.
\newblock {\em The New York Times}, 26, 2016.

\bibitem{Matko2019}
K.~Matko and P.~Sedlmeier.
\newblock What is meditation? proposing an empirically derived classification
  system.
\newblock {\em Frontiers in psychology}, 10:2276, 2019.

\bibitem{matzer2018combining}
F.~Matzer, E.~Nagele, N.~Lerch, C.~Vajda, and C.~Fazekas.
\newblock Combining walking and relaxation for stress reduction—a randomized
  cross-over trial in healthy adults.
\newblock {\em Stress and Health}, 34(2):266--277, 2018.

\bibitem{mccann2014virtual}
R.~A. McCann, C.~M. Armstrong, N.~A. Skopp, A.~Edwards-Stewart, D.~J.
  Smolenski, J.~D. June, M.~Metzger-Abamukong, and G.~M. Reger.
\newblock Virtual reality exposure therapy for the treatment of anxiety
  disorders: an evaluation of research quality.
\newblock {\em Journal of anxiety disorders}, 28(6):625--631, 2014.

\bibitem{menelas2018use}
B.-A.~J. Menelas, C.~Haidon, A.~Ecrepont, and B.~Girard.
\newblock Use of virtual reality technologies as an action-cue exposure therapy
  for truck drivers suffering from post-traumatic stress disorder.
\newblock {\em Entertainment computing}, 24:1--9, 2018.

\bibitem{meyns2017effect}
P.~Meyns, L.~Pans, K.~Plasmans, L.~Heyrman, K.~Desloovere, and G.~Molenaers.
\newblock The effect of additional virtual reality training on balance in
  children with cerebral palsy after lower limb surgery: A feasibility study.
\newblock {\em Games for Health Journal}, 6(1):39--48, 2017.

\bibitem{mistry2020meditating}
D.~Mistry, J.~Zhu, P.~Tremblay, C.~Wekerle, R.~Lanius, R.~Jetly, and P.~Frewen.
\newblock Meditating in virtual reality: Proof-of-concept intervention for
  posttraumatic stress.
\newblock {\em Psychological Trauma: Theory, Research, Practice, and Policy},
  12(8):847, 2020.

\bibitem{MONROE2016109}
S.~Monroe and G.~Slavich.
\newblock Psychological stressors: Overview.
\newblock In G.~Fink, ed., {\em Stress: Concepts, Cognition, Emotion, and
  Behavior}, pp. 109--115. Academic Press, San Diego, 2016. doi: {{%
10\hspace{.1pt}\discretionary{.}{%
}{.}\hspace{.4pt}1016\discretionary{/}{%
}{/}B978\discretionary{%
}{-}{-}0\discretionary{%
}{-}{-}12\discretionary{%
}{-}{-}800951\discretionary{%
}{-}{-}2\hspace{.1pt}\discretionary{.}{%
}{.}\hspace{.4pt}00013\discretionary{%
}{-}{-}3}}


\bibitem{montero2015mindfulness}
J.~Montero-Marin, M.~Tops, R.~Manzanera, M.~M. Piva~Demarzo, M.~Alvarez~de Mon,
  and J.~Garc{\'\i}a-Campayo.
\newblock Mindfulness, resilience, and burnout subtypes in primary care
  physicians: The possible mediating role of positive and negative affect.
\newblock {\em Frontiers in psychology}, 6:1895, 2015.

\bibitem{nordh2009components}
H.~Nordh, T.~Hartig, C.~Hagerhall, and G.~Fry.
\newblock Components of small urban parks that predict the possibility for
  restoration.
\newblock {\em Urban forestry \& urban greening}, 8(4):225--235, 2009.

\bibitem{nordh2013pocket}
H.~Nordh and K.~{\O}stby.
\newblock Pocket parks for people--a study of park design and use.
\newblock {\em Urban forestry \& urban greening}, 12(1):12--17, 2013.

\bibitem{Oman2008}
D.~Oman, S.~L. Shapiro, C.~E. Thoresen, T.~G. Plante, and T.~Flinders.
\newblock Meditation lowers stress and supports forgiveness among college
  students: A randomized controlled trial.
\newblock {\em Journal of American College Health}, 56(5):569--578, 2008. doi:
  {{%
10\hspace{.1pt}\discretionary{.}{%
}{.}\hspace{.4pt}3200\discretionary{/}{%
}{/}JACH\hspace{.1pt}\discretionary{.}{%
}{.}\hspace{.4pt}56\hspace{.1pt}\discretionary{.}{%
}{.}\hspace{.4pt}5\hspace{.1pt}\discretionary{.}{%
}{.}\hspace{.4pt}569\discretionary{%
}{-}{-}578}}


\bibitem{opdyke1995effectiveness}
D.~Opdyke, J.~S. Williford, and M.~North.
\newblock Effectiveness of computer-generated (virtual reality) graded exposure
  in the treatment of acrophobia.
\newblock {\em Am J psychiatry}, 1(152):626--628, 1995.

\bibitem{ospina2007meditation}
M.~B. Ospina, K.~Bond, M.~Karkhaneh, L.~Tjosvold, B.~Vandermeer, Y.~Liang,
  L.~Bialy, N.~Hooton, N.~Buscemi, D.~M. Dryden, et~al.
\newblock Meditation practices for health: state of the research.
\newblock {\em Evidence report/technology assessment}, (155):1--263, 2007.

\bibitem{park2019literature}
M.~J. Park, D.~J. Kim, U.~Lee, E.~J. Na, and H.~J. Jeon.
\newblock A literature overview of virtual reality (vr) in treatment of
  psychiatric disorders: recent advances and limitations.
\newblock {\em Frontiers in psychiatry}, 10:505, 2019.

\bibitem{persson2021virtual}
J.~Persson, D.~Clifford, M.~Wallerg{\aa}rd, U.~Sand{\'e}n, et~al.
\newblock A virtual smash room for venting frustration or just having fun:
  Participatory design of virtual environments in digitally reinforced cancer
  rehabilitation.
\newblock {\em JMIR Rehabilitation and Assistive Technologies}, 8(4):e29763,
  2021.

\bibitem{qian2020effectiveness}
J.~Qian, D.~J. McDonough, and Z.~Gao.
\newblock The effectiveness of virtual reality exercise on individual’s
  physiological, psychological and rehabilitative outcomes: A systematic
  review.
\newblock {\em International journal of environmental research and public
  health}, 17(11):4133, 2020.

\bibitem{reger2019does}
G.~M. Reger, D.~Smolenski, A.~Edwards-Stewart, N.~A. Skopp, A.~S. Rizzo, and
  A.~Norr.
\newblock Does virtual reality increase simulator sickness during exposure
  therapy for post-traumatic stress disorder?
\newblock {\em Telemedicine and e-Health}, 25(9):859--861, 2019.

\bibitem{Blender}
\relax{Blender Foundation}.
\newblock Blender.
\newblock Accessed: 2023-3-24.

\bibitem{Salleh2008}
M.~R. Salleh.
\newblock Life event, stress and illness.
\newblock {\em The Malaysian journal of medical sciences: MJMS}, 15(4):9, 2008.

\bibitem{sedlmeier2018psychological}
P.~Sedlmeier, C.~Lo{\ss}e, and L.~C. Quasten.
\newblock Psychological effects of meditation for healthy practitioners: an
  update.
\newblock {\em Mindfulness}, 9(2):371--387, 2018.

\bibitem{Southwick2016}
S.~M. Southwick, L.~Sippel, J.~Krystal, D.~Charney, L.~Mayes, and R.~Pietrzak.
\newblock {Why are some individuals more resilient than others: the role of
  social support}.
\newblock {\em World Psychiatry}, 15(1):77--79, 2016. doi: {{%
10\hspace{.1pt}\discretionary{.}{%
}{.}\hspace{.4pt}1002\discretionary{/}{%
}{/}wps\hspace{.1pt}\discretionary{.}{%
}{.}\hspace{.4pt}20282}}


\bibitem{spielberger1983state}
C.~D. Spielberger.
\newblock State-trait anxiety inventory for adults.
\newblock 1983.

\bibitem{Spielberger1971}
C.~D. Spielberger, F.~Gonzalez-Reigosa, A.~Martinez-Urrutia, L.~F. Natalicio,
  and D.~S. Natalicio.
\newblock The state-trait anxiety inventory.
\newblock {\em Revista Interamericana de Psicologia/Interamerican Journal of
  Psychology}, 5(3 \& 4), 1971.

\bibitem{tarrant2022feasibility}
J.~Tarrant, R.~Jackson, and J.~Viczko.
\newblock A feasibility test of a brief mobile virtual reality meditation for
  frontline healthcare workers in a hospital setting. front.
\newblock {\em Virtual Real. 3: 764745. doi: 10.3389/frvir}, 2022.

\bibitem{tarrant2018virtual}
J.~Tarrant, J.~Viczko, and H.~Cope.
\newblock Virtual reality for anxiety reduction demonstrated by quantitative
  eeg: a pilot study.
\newblock {\em Frontiers in psychology}, 9:1280, 2018.

\bibitem{tombaugh2006comprehensive}
T.~N. Tombaugh.
\newblock A comprehensive review of the paced auditory serial addition test
  (pasat).
\newblock {\em Archives of clinical neuropsychology}, 21(1):53--76, 2006.

\bibitem{tosevski2010personality}
D.~L. Tosevski, M.~P. Milovancevic, and S.~D. Gajic.
\newblock Personality and psychopathology of university students.
\newblock {\em Current opinion in psychiatry}, 23(1):48--52, 2010.

\bibitem{vaquero2020virtual}
M.~A. Vaquero-Blasco, E.~Perez-Valero, M.~A. Lopez-Gordo, and C.~Morillas.
\newblock Virtual reality as a portable alternative to chromotherapy rooms for
  stress relief: a preliminary study.
\newblock {\em Sensors}, 20(21):6211, 2020.

\bibitem{vaquero2021virtual}
M.~A. Vaquero-Blasco, E.~Perez-Valero, C.~Morillas, and M.~A. Lopez-Gordo.
\newblock Virtual reality customized 360-degree experiences for stress relief.
\newblock {\em Sensors}, 21(6):2219, 2021.

\bibitem{waller2021meditating}
M.~Waller, D.~Mistry, R.~Jetly, and P.~Frewen.
\newblock Meditating in virtual reality 3: 360 video of perceptual presence of
  instructor.
\newblock {\em Mindfulness}, 12(6):1424--1437, 2021.

\bibitem{weckesser2019psychometric}
L.~J. Weckesser, F.~Dietz, K.~Schmidt, J.~Grass, C.~Kirschbaum, and R.~Miller.
\newblock The psychometric properties and temporal dynamics of subjective
  stress, retrospectively assessed by different informants and questionnaires,
  and hair cortisol concentrations.
\newblock {\em Scientific Reports}, 9(1):1--12, 2019.

\bibitem{xie2009clinical}
H.~Xie, J.~Kang, and G.~H. Mills.
\newblock Clinical review: The impact of noise on patients' sleep and the
  effectiveness of noise reduction strategies in intensive care units.
\newblock {\em Critical Care}, 13(2):1--8, 2009.

\bibitem{yang2018happier}
E.~Yang, E.~Schamber, R.~M. Meyer, and J.~I. Gold.
\newblock Happier healers: randomized controlled trial of mobile mindfulness
  for stress management.
\newblock {\em The Journal of Alternative and Complementary Medicine},
  24(5):505--513, 2018.

\bibitem{Yaribeygi2017}
H.~Yaribeygi, Y.~Panahi, H.~Sahraei, T.~P. Johnston, and A.~Sahebkar.
\newblock The impact of stress on body function: a review.
\newblock {\em EXCLI Journal; 16:Doc1057; ISSN 1611-2156}, 2017.
\newblock Publisher: IfADo - Leibniz Research Centre for Working Environment
  and Human Factors, Dortmund. doi: {{%
10\hspace{.1pt}\discretionary{.}{%
}{.}\hspace{.4pt}17179\discretionary{/}{%
}{/}EXCLI2017\discretionary{%
}{-}{-}480}}


\bibitem{yildirim2020efficacy}
C.~Yildirim and T.~O’Grady.
\newblock The efficacy of a virtual reality-based mindfulness intervention.
\newblock In {\em 2020 IEEE International Conference on Artificial Intelligence
  and Virtual Reality (AIVR)}, pp. 158--165. IEEE, 2020.

\bibitem{zaharuddin2019factors}
F.~A. Zaharuddin, N.~Ibrahim, M.~E. Rusli, E.~M. Mohd~Mahidin, and A.~M. Yusof.
\newblock Factors to consider when designing a virtual environment to treat
  stress.
\newblock In {\em International Visual Informatics Conference}, pp. 36--47.
  Springer, 2019.

\bibitem{zainudin2014stress}
A.~R.~R. Zainudin, A.~M. Yusof, M.~E. Rusli, M.~Z.~M. Yusof, and I.~Mahalil.
\newblock Stress treatment: The effectiveness between guided and non-guided
  virtual reality setup.
\newblock In {\em Proceedings of the 6th International Conference on
  Information Technology and Multimedia}, pp. 374--379. IEEE, 2014.

\bibitem{zillig2002we}
L.~M.~P. Zillig, S.~H. Hemenover, and R.~A. Dienstbier.
\newblock What do we assess when we assess a big 5 trait? a content analysis of
  the affective, behavioral, and cognitive processes represented in big 5
  personality inventories.
\newblock {\em Personality and Social Psychology Bulletin}, 28(6):847--858,
  2002.

\end{thebibliography}
\end{document}